\documentstyle[aps,pra]{revtex}

\begin{document}
\title{Elementary operations for quantum logic with a single trapped two-level 
cold ion beyond Lamb-Dicke limit}
\author{L. F. Wei, S. Y. Liu and X. L. Lei}
\address{Department of Physics, Shanghai Jiao Tong University,
1954 Huashan Road, Shanghai 200030, China}

\maketitle

\begin{abstract}
A simple alternative scheme for implementing quantum gates with a single trapped cold two-level ion 
beyond the Lamb-Dicke (LD) limit is proposed. Basing on the quantum dynamics for the
laser-ion interaction described by a generalized Jaynes-Cummings model,
one can introduce two kinds of elementary quantum operations i.e.,  
the simple rotation on the bare atomic state,  
generated by applying a resonant pulse, and 
the joint operation on the internal and external degrees of the ion, 
performed by using an off-resonant pulse. Several typical quantum 
gates, including Hadamard gate, controlled-Z and controlled-NOT gates 
$etc.$, can thus be implemented exactly by 
using these elementary operations. The experimental 
parameters including the LD parameter and
the durations of the applied laser pulses, for these implementation are derived 
analytically and numerically.
Neither the LD approximation for the laser-ion interaction nor the
auxiliary atomic level is needed in the present scheme.
 
\vspace{0.2cm}
PACS number(s): 03.67.Lx; 42.50. Dv; 03.65.Bz
\end{abstract}
\vspace{0.2cm}
\section{introduction}

Since Shor's algorithm for efficiently factoring large numbers
was proposed \cite{Shor}, many studies have been carried out with a view
to implement the quantum computation \cite{EJ}.
It has been shown that a simple rotation in the Hilbert space of
an individual qubit, i.e. one-qubit gate,
and a controlled rotation, such as a controlled-NOT ($CN$) or
controlled-Z ($CZ$), between two different qubits,
i.e. two-qubit controlled gate, are fundamental quantum gates.
Any unitary transformation on arbitrarily many computational qubits
can be carried out by
repeatedly performing these fundamental gates \cite{Bar}.
Several kinds of physical systems, e.g. nuclear magnetic resonance (NMR) 
\cite{GC}, coupled quantum dots \cite{LD} and the cavity QED \cite{Tru} $etc.$,
have been proposed to implement these fundamental quantum gates.
Quantum computation with trapped cold ions, introduced first by Cirac and Zoller\cite{CZ},
have been paid much attention theoretically and experimentally \cite{Ste}. 
This scheme is based on the laser-ion
interaction and possesses a long qubit coherence times \cite{KS}.
Information is stored in the spin states of an array of trapped cold ions
and manipulated by using laser pulses. In 1995, the fundamental quantum gates with
a single trapped cold ion had been demonstrated experimentally \cite{Mor1}.  
In this scheme the ground and first excited states of the 
external vibration of the ion are used to encode the control qubit and the target 
qubit are encoded by two states
of the internal degree freedom of the ion. However, the experimental operation  
involves an auxiliary atomic level and is limited in LD regime. 
Without performing the LD approximation and needing any auxiliary atomic level,
some authors \cite{Mor2} \cite{LG} proposed another alternative method to realize 
another controlled operation, which is equivalent to the exact $CN$ logic 
operation \cite{Mor1} apart from phase factors, in a single trapped two-level cold ion
by applying a single resonant pulse. 
In a recent work we proved that the exact $CN$ logic 
operation between the internal and external degrees of freedom of the ion  
can be realized assuredly by using a single off-resonant laser pulse, if the target 
qubit is encoded by the dressed atomic states \cite{wei}. 
Adopting the usual encode scheme for conveniently measuring, i.e. the target 
qubit is encoded by the bare atomic states ($|g\rangle$ and $|e\rangle$),
we show in this paper that the fundamental quantum gates e.g. the Hadamard ($H$) gate and the controlled-Z ($CZ$) or $CN$ gate, can be implemented exactly 
in a single trapped ion register by 
using two kinds of elementary operations generated by 
applying different frequencies laser pulses. The experimental parameters for these
realizations are derived by analytically and numerically.   
The multiquantum 
interaction between the ion and the applied laser is considered in the present work.  

\section{Elementary operations for manipulating quantum state of laser-ion system beyond 
Lamb-Dicke limit}

For simplicity, we assume that a single ion is stored in a coaxial
resonator RF-ion trap\cite {Jef}, which provides pseudopotential oscillation
frequencies satisfying the condition $\omega _{x}<<\omega _{y,z}$ along
the principal axes of the trap. Only the quantized vibrational motion
along $x$ direction is considered for the cooled ion\cite{Mor1}.
Following Blockley {\em et al }\cite{Blo}, the
interaction between a single trapped two-level cold ion and a classical
single-mode travelling light field of frequency $\omega _{L}$ can be
described by the following Hamiltonian:
\begin{eqnarray}
\hat{H}(t)=\hbar \omega (\hat{a}^{\dagger }\hat{a}+\frac{1}{2})+\frac{1}{2}%
\hbar \omega _{0}\hat{\sigma}_{z}+\frac{\hbar \Omega }{2}\{\hat{\sigma}%
_{+}\exp [i\eta (\hat{a}+\hat{a}^{\dagger })-i(\omega _{L}t+\phi )]+h.c.\}.
\end{eqnarray}
The first two terms correspond to the ion's kinetic and potential energy in
the trap respectively, $\omega $ being the trap frequency.
The final term gives the interaction between the ion and the light field with phase $\phi $.
Pauli operators $\hat{\sigma}_{z}$ and $\hat{\sigma}_{\pm }$ describing the
internal degrees of freedom of the ion. $\hat{a}^{\dagger }$ and $\hat{a}$
are the creation and annihilation operators of the trap vibrational quanta.
$\omega _{0}$ is the atomic transition frequency, $\Omega$ is the
Rabi frequency. $\eta (<1)$ is the LD parameter. Without loss of
generality and for simplicity, we assumed that the laser is resonant with the
$k$th red-shifted vibrational sideband i.e. the frequency of laser field
is chosen as $\omega_{L}=\omega _{0}-k\omega $. Under the usual
rotating-wave approximation, the Hamiltonian of the system reads
\begin{eqnarray}
\hat{H}=\frac{\hbar \Omega }{2}\exp (-\frac{\eta ^{2}}{2}-i\phi)\{\hat{\sigma}%
_{+}(i\eta )^{k}[\sum_{n=0}^{\infty }\frac{(i\eta )^{2n}\hat{a}^{\dagger n}%
\hat{a}^{n}}{n!(n+k)!}]\hat{a}^{k}+h.c.\},
\end{eqnarray}
in the interaction picture defined by the unitary operator $\hat{U}%
_{0}(t)=\exp [-i\omega t(\hat{a}^{\dagger }\hat{a}+1/2)]\exp (-it\delta
\hat{\sigma }_{z}/2)$. Where $\delta =\omega _{0}-\omega _{L}$ is the
detuning of laser field with the ion. The above Hamiltonian is similar to
that of nonlinear coupled multiquantum Jaynes-Cummings model \cite{Vog},
which is exactly solvable. Therefore, it is easy to check the dynamical
evolution of any two-qubit initial state by using evolution operator
$\hat{U}(t)=\exp(-\frac{i}{\hbar }\hat{H}t)$.
Indeed, with the help of relation \cite{Ste}\cite{Win} 
$$
\langle m-k|\langle e|\hat{H}|m\rangle|g\rangle=\left\{
\begin{array}{l}
0 ,\qquad m<k; \\
\\
\\
\hbar i^k e^{-i\phi}\Omega _{m-k,m},\qquad m\geq k,
\end{array}
\right.
$$
with 
$$
\Omega _{m-k,m}=\frac{\Omega \eta ^{k}e^{-\eta ^{2}/2}}{2}\sqrt{\frac{(m)!}{(m-k)!}%
}\sum_{n=0}^{m-k}\frac{(i\eta)^{2n}}{(k+n)!}C^n_{m-k},
$$
the time evolution of the initial
states $|m\rangle |e\rangle $ and $|m\rangle|g\rangle$ can be expressed as
\begin{eqnarray}
|m\rangle |e\rangle \longrightarrow \cos \Omega _{m,m+k}t|m\rangle |e\rangle
-(-i)^{k-1}e^{i\phi }\sin \Omega _{m,m+k} t|m+k\rangle |g\rangle ,
\end{eqnarray}
and
\begin{eqnarray}
|m\rangle |g\rangle \longrightarrow \left\{
\begin{array}{l}
|m\rangle |g\rangle ,\qquad m<k;\vspace{0.2cm} \\
\\
\cos \Omega _{m-k,m}t|m\rangle |g\rangle +i^{k-1}e^{-i\phi }\sin \Omega
_{m-k,m}t|m-k\rangle |e\rangle ,\qquad m\geq k,
\end{array}
\right.
\end{eqnarray}
respectively. The above treatment can also be modified directly to another
laser excitation case, i.e., the $k$th blue-sideband laser addresses on the
specifically chosen ion. In the present work only the red-sideband excitation is considered.
It is seen from equations (3) and (4) that entangled states are caused from
the time evolution of the state $|m\rangle |e\rangle $ and the state
$|m\rangle |g\rangle $ with $m\geq k.$ From this conditional quantum dynamics we can 
define two kinds of basic quantum operations: one 
is the simple rotations on the target qubit,  
\begin{eqnarray}
\hat{r}_m(\phi,t)=\{\cos \Omega _{m,m}t|g\rangle \langle g|-ie^{-i\phi }\sin \Omega _{m,m}t
|e\rangle \langle g|-ie^{i\phi }\sin \Omega _{m,m}t|g\rangle \langle
e|+\cos \Omega _{m,m}t|e\rangle \langle e|\}\otimes |m\rangle\langle m|,
\end{eqnarray}
generated by applying a resonant pulse ($\omega_L=\omega_0$) to the ion, and 
another is the two-qubit joint operation 
\begin{eqnarray}
\hat{R}_m(\phi,t)=\left\{
\begin{array}{ll}
|m\rangle|g\rangle\langle m|\langle g|+
(\cos\Omega _{m,m+k}t|m\rangle|e\rangle-
(-i)^{k-1}e^{i\phi}\sin\Omega _{m,m+k}t|m+k\rangle|g\rangle)\langle m|\langle e|, \hspace{0.2cm}m<k;\\
\\
\\
(\cos\Omega_{m-k,m}t|m\rangle|g\rangle+
i^{k-1}e^{-i\phi}\sin\Omega _{m-k,m}t|m-k\rangle|e\rangle)\langle m|\langle g|\\
\\
\hspace{2.7cm}+(\cos\Omega _{m,m+k}t|m\rangle|e\rangle-
(-i)^{k-1}e^{i\phi}\sin\Omega _{m,m+k}t|m+k\rangle|g\rangle)\langle m|\langle e|, \hspace{0.2cm}m\geq k,
\end{array}
\right.
\end{eqnarray}  
performed by using an off-resonant pulse. Here $\phi$ and $t$ are the
initial phase and duration of the applied pulse, respectively. 
We note that the state of the control qubit
is unchanged during the operation $\hat{r}_m(\phi,t)$.
It is worthwhile to point out that the qubit 
encoded by two Fock states of the external vibration of the ion can only be regarded as 
the control qubit, as all Fock states of the external vibration 
are present simultaneously. As a consequence, only the
qubit encoded by two internal states of the ion can be serviced
as the target qubit. Therefore, in what follows we only discuss how to implement 
the quantum operation on the internal state of the ion.  
 
\section{Quantum gates with a single trapped two-level cold ion}

As we well know that computation on a register of qubits can be broken up into a series of two-qubit operations, accompanied by one-qubit rotations \cite{Bar}. For the present system the operations $\hat{r}_m(\phi,t)$ and $\hat{R}_m(\phi,t)$ thus form a universal set i.e. any unitary operation on the internal state of the ion can be carried out by using the elements of the set repeatedly.
Without the loss of generality, we now show how to implement three typical quantum 
logic operations : 1). Hadamard ($H$) gate, 2). controlled-Z ($CZ$) gate and  3). controlled-NOT ($CN$) gate, respectively, in this simple quantum register. 

Firstly, we discuss how to realize the $H$ gate operation 
\begin{equation}
\hat{H}_m=\frac{1}{\sqrt{2}}\{|g\rangle\langle g|+|e\rangle \langle g|
+|g\rangle \langle e|-|e\rangle\langle e|\}\otimes |m\rangle\langle m|.
\end{equation} 
It generates the uniform superposition of two encoded states of the target qubit from 
one of them. The control qubit remains in its initial state after this operation. The $H$ gate takes an important role in quantum computation, as almost all initial states of quantum computing are the uniform superposition of the encoded states
the qubits. One can easily see from the discussion in Sec. II that the $H$ gate operation (7) can not be carried out exactly, if only the resonant laser pulse is applied to the ion. We now consider the  
combination operation $\hat{R}_m(\phi_2,t_2)\hat{r}_m(\phi_1,t_1)$ with $m <k $,
It is performed by sequentially applying a resonant pulse and an off-resonant pulse. 
Under this operation the evolution of the initial states $|m\rangle|g\rangle$ and $|m\rangle|e\rangle$ can be expressed as      
\begin{eqnarray}
|m\rangle|g\rangle&\longrightarrow&
\hat{R}_m(\phi_2,t_2)\hat{r}_m(\phi_1,t_1)|m\rangle|g\rangle \nonumber\\
\\ \nonumber
&=&\cos\Omega_{m,m}t_1|m\rangle|g\rangle
-ie^{-i\phi_1}\sin\Omega_{m,m}t_1
(\cos\Omega_{m,m+k}t_2|m\rangle|e\rangle
-(-i)^{k-1}e^{i\phi_2}\sin\Omega_{m,m+k}t_2|m+k\rangle|g\rangle),
\end{eqnarray}
and 
\begin{eqnarray}
|m\rangle|e\rangle&\longrightarrow&
\hat{R}_m(\phi_2,t_2)\hat{r}_m(\phi_1,t_1)|m\rangle|e\rangle \nonumber\\
\\ \nonumber
&=&\cos\Omega_{m,m}t_1
(\cos\Omega_{m,m+k}t_2|m\rangle|e\rangle
-(-i)^{k-1}e^{i\phi_2}\sin\Omega_{m,m+k}t_2|m+k\rangle|g\rangle)
-ie^{i\phi_1}\sin\Omega_{m,m}t_1|m\rangle|g\rangle,
\end{eqnarray}
respectively. Obviously, if the following matching conditions 
\begin{equation}
\cos\Omega_{m,m}t_1=\frac{1}{\sqrt{2}},\hspace{3mm}
\cos\Omega_{m,m+k}t_2=-1,\hspace{3mm}\phi_1=\pi/2\pm 2k\pi, k=1,2,...,
\end{equation}
are satisfied, the $H$ gate (7) is realized exactly. This mean that the exact $H$ gate operation can be 
implemented by sequentially applying a resonant pulse and an off-resonant pulse,
i.e.,
\begin{equation}
\hat{R}_m(T_f)\hat{r}_m(\phi_r,T_r)\longrightarrow \hat{H}_m,
\end{equation}
with
$$
T_f=(\pi\pm 2k\pi)/\Omega_{m,m+k}, \hspace{0.4cm}
T_r=(\pi/4\pm 2k\pi)/\Omega_{m,m}, \hspace{0.4cm}\phi_r=\pi/2\pm 2k\pi.
$$
Similarly, one can prove also that 
\begin{equation}
\hat{r}_m(3\pi/2\pm 2k\pi,T_r)\hat{R}_m(T_f)\longrightarrow\hat{H}_m.  
\end{equation}
Therefore, the exact $H$ gate operation on the internal state of the ion can be 
realized by applying a resonant and an off-resonant pulses sequentially. 

Secondly, we want to realize the typical two-qubit controlled gate e.g. the controlled-Z ($CZ$) gate, which takes the form
\begin{eqnarray}
\hat{CZ}_{01}=|0\rangle |g\rangle\langle 0|\langle g|+|0\rangle |e\rangle\langle 0|\langle e|
+|1\rangle |g\rangle\langle 1|\langle g|-|1\rangle |e\rangle\langle 1|\langle e|
=\left(
\begin{array}{cccc}
1 & 0 & 0 & 0 \\
0 & 1 & 0 & 0 \\
0 & 0 & 1 & 0 \\
0 & 0 & 0 & -1
\end{array}
\right),
\end{eqnarray}
in the computational 
space $\Gamma:\Gamma=\{|0\rangle,|1\rangle\} \otimes \{|g\rangle,|e\rangle\}$.    
The control qubit in the present work is encoded by $|0\rangle$ and $|1\rangle$, i.e. the ground and first 
excited states of the external vibration of the ion. This gate is called also the controlled-rotation ($CR$) operation, which represents that the state of the target 
qubit is rotated by the Pauli operator $\sigma_z$ if and only if the control qubit is in the state $|1\rangle$.
We now show that the exact $\hat{CZ}_{01}$ gate can be implemented exactly by a single off-resonant pulse, 
once the experimental parameters e.g. the 
LD parameter $\eta$ and duration of the applied pulse. In the computational 
space $\Gamma$ the two-qubit operation 
$\hat{R}_m (\phi,t)$, performed by using a $k$th red-sideband pulse with initial phase $\phi$ and duration $t$,  
takes the following form
\begin{eqnarray}
\hat{R}_{01}(\phi,t)&=&|0\rangle|g\rangle\langle 0|\langle g|
+(\cos\Omega_{0,1}t|0\rangle|e\rangle
-e^{i\phi}\sin\Omega_{0,1}t|1\rangle|g\rangle)
\langle 0|\langle e|\\\nonumber
\\\nonumber
&+&(\cos\Omega_{0,1}|1\rangle|g\rangle+e^{-i\phi}\sin\Omega_{0,1}t|0\rangle|e\rangle)
\langle 1|\langle g|
+(\cos\Omega_{1,2}t|1\rangle|e\rangle
-e^{i\phi}\sin\Omega_{1,2}t|2\rangle|g\rangle)\langle 1|\langle e|,
\end{eqnarray} 
for $k=1$, and 
\begin{eqnarray}
\hat{R}_{01}(\phi,t)&=&|0\rangle|g\rangle\langle 0|\langle g|
+(\cos\Omega_{0,k}t|0\rangle|e\rangle-
(-i)^{k-1}e^{i\phi}\sin\Omega_{0,k}t|k\rangle|g\rangle)
\langle 0|\langle e|\\\nonumber
\\\nonumber
&+&|1\rangle|g\rangle\langle 1|\langle g|
+(\cos\Omega_{1,k+1}t|1\rangle|e\rangle-
(-i)^{k-1}e^{i\phi}\sin\Omega_{1,k+1}t|k+1\rangle|g\rangle)
\langle 1|\langle e|,
\end{eqnarray}
for $k\geq 2$. Obviously, if the LD parameter $\eta$ and the duration of the applied 
off-resonant pulse are set up properly to satisfy the condition
\begin{equation}
\cos \Omega _{0,k}t=1,\hspace{1cm}\cos \Omega _{1,k+1}t=-1,
\end{equation}
the two-qubit operation $\hat{R}_{01}$ reduces to the exact $CZ$ logic gate. 
It is worth noting that there is no any requirement on the
initial phase of the applied pulse for realizing
$CZ$ gate (13).

Finally, we turn to realize the exact controlled-NOT 
($CN$) logic gate with a single trapped two-level cold ion. This gate takes the form  
\begin{equation}
\hat{CN}_{01}=|0\rangle |g\rangle\langle 0|\langle g|+|0\rangle |e\rangle\langle 0|\langle e|
+|1\rangle |e\rangle\langle 1|\langle g|+|1\rangle |g\rangle\langle 1|\langle e|
=\left( 
\begin{array}{cccc}
1& 0 & 0 & 0 \\ 
0 & 1 & 0 & 0 \\ 
0 & 0 & 0 & 1 \\ 
0 & 0 & 1 & 0
\end{array}
\right)
\end{equation}   
in the $\Gamma$ space defined above. It can also be called as controlled-X ($CX$) gate, 
as it represents that the target qubit is rotated by Pauli operator
$\sigma_x$ when the control qubit is in the state $|1\rangle$. 
Liking to the recent works \cite{Mor2}\cite{LG}, 
we can easily show that a single resonant pulse may also result in 
a two-qubit controlled operation, i.e., a resonant pulse satisfying the condition  
$$
\cos \Omega _{0,0}t=1, \hspace{3mm} \sin\Omega _{1,1}t=1,
$$
yields the controlled operation \cite{Mor2,LG}
$$
\hat{r}_{01}(\phi)=\left(
\begin{array}{cccc}
1 & 0 & 0 & 0 \\
0 & 1 & 0 & 0 \\
0 & 0 & 0 & -ie^{i\phi }\\
0 & 0 & -ie^{-i\phi }& 0
\end{array}
\right),
$$
which is equivalent to the exact $CN$ gate \cite{Mor1}, apart from 
phase factors that must be eliminated by applying subsequent operations.
In what follows we show how to implement the exact $CN$ gate with a single trapped 
two-level cold ion beyond LD limit by using three pulse sequentially.   
Indeed, one can easily prove that  
\begin{eqnarray}
\hat{r}_{01}(\phi_3,t_3)\hat{CZ}_{01}\hat{r}_{01}(\phi_1,t_1)=\left( 
\begin{array}{cccc}
A_{0000}& A_{0001} & 0        & 0      \\ 
A_{0100}& A_{0101} & 0        & 0      \\ 
0       & 0        & A_{1000} & A_{1011}\\ 
0       & 0        & A_{1110} & A_{1111} 
\end{array}
\right ),
\end{eqnarray}
with
\begin{eqnarray}
\begin{array}{l}
A_{0000}=\cos\Omega_{0,0}t_3\cos\Omega_{0,0}t_1-e^{i(\phi_3-\phi_1)}\sin\Omega_{0,0}t_3
\sin\Omega_{0,0}t_1 ,\\
\vspace{0.2cm}
A_{0001}=-ie^{i\phi_1}\cos\Omega_{0,0}t_3\sin\Omega_{0,0}t_1-ie^{i\phi_3}
\sin\Omega_{0,0}t_3\cos\Omega_{0,0}t_1,\\
\vspace{0.2cm}
A_{0100}=-ie^{-i\phi_3}\sin\Omega_{0,0}t_3\cos\Omega_{0,0}t_1-ie^{-i\phi_1}
\cos\Omega_{0,0}t_3\sin\Omega_{0,0}t_1,\\
\vspace{0.2cm}
A_{0101}=\cos\Omega_{0,0}t_3\cos\Omega_{0,0}t_1-e^{-i(\phi_3-\phi_1)}
\sin\Omega_{0,0}t_3\sin\Omega_{0,0}t_1;\\
\vspace{0.2cm}
A_{1010}=\cos\Omega_{1,1}t_3\cos\Omega_{1,1}t_1+e^{i(\phi_3-\phi_1)}
\sin\Omega_{1,1}t_3\sin\Omega_{1,1}t_1,\\
\vspace{0.2cm}
A_{1011}=-ie^{i\phi_1}\cos\Omega_{1,1}t_3\sin\Omega_{1,1}t_1+ie^{i\phi_3}
\sin\Omega_{1,1}t_3\cos\Omega_{1,1}t_1,\\
\vspace{0.2cm}
A_{1110}=-ie^{i\phi_1}\cos\Omega_{1,1}t_3\sin\Omega_{1,1}t_1+ie^{i\phi_3}
\sin\Omega_{1,1}t_3\cos\Omega_{1,1}t_1,\\
\vspace{0.2cm}
A_{1111}=-\cos\Omega_{1,1}t_3\cos\Omega_{1,1}t_1-e^{-i(\phi_3-\phi_1)}
\sin\Omega_{1,1}t_3\sin\Omega_{1,1}t_1,
\end{array}
\end{eqnarray}
with $t_1$ and $t_3$ the durations of the first and third applied resonant 
laser pulses, respectively. 
If two initial phases $\phi_1,\phi_3$ satisfy the relation
\begin{equation}
\phi_3-\phi_1=\pm 2k\pi,\hspace{0.2cm}k=0,1,2,...,
\end{equation}
the matrix elements on the right side of Eq.\,(19) become
$$
\left\{
\begin{array}{l}
A_{0000}=A_{0101}=\cos \Omega _{0,0}(t_{3}+t_1),\hspace{2mm}
A_{0001}=-ie^{i\phi _{1}}\sin \Omega _{0,0}(t_{3}+t_1),\hspace{0.2cm}
A_{0100}=-ie^{-i\phi _{1}}\sin \Omega _{0,0}(t_{3}+t_1),\\
\\
A_{0101}=-A_{1111}=\cos \Omega _{0,0}(t_{3}-t_1),\hspace{2mm}
A_{1011}=ie^{i\phi _{1}}\sin \Omega _{1,1}(t_3-t_{1}),\hspace{0.2cm}
A_{1110}=-ie^{-i\phi _{1}}\sin \Omega _{1,1}(t_3-t_{1}).
\end{array}
\right.
$$
Furthermore, if the following matching conditions
\begin{eqnarray}
\cos \Omega _{0,0}(t_{3}+t_{1})=1,\hspace{0.3cm}
\sin \Omega _{1,1}(t_{3}-t_{1})=\left\{
\begin{array}{ll}
1 & \mbox{for $\phi_1=3\pi/2+2k'\pi, k'=0,1,2,...;$} \\
\\
-1 & \mbox{for $\phi=\pi/2+2k'\pi, k=0,1,2,...,$}
\end{array}
\right .
\end{eqnarray}
are satisfied, we have 
$$
A_{0000}=A_{0101}=A_{1011}=A_{1110}=1,\hspace{0.3cm}
A_{0001}=A_{0100}=A_{1000}=A_{1111}=0.
$$
This means that, under the conditions (20) and (21), the operation (18) 
reduces to the exact reduced $CN$ logic gate \cite{Mor1}, i.e.,
\begin{eqnarray}
\hat{r}_{01}(\phi_3,t_{3})\hat{CZ}_{01}\hat{r}_{01}(\phi_1,t_{1})
\longrightarrow \left(
\begin{array}{cccc}
1 & 0 & 0 & 0 \\
0 & 1 & 0 & 0 \\
0 & 0 & 0 & 1 \\
0 & 0 & 1 & 0
\end{array}
\right) =\hat{CN}_{01}.
\end{eqnarray}

Specifically, we now discuss how to set up the experimental parameters 
to realize the $\hat{CN}$ logic operation (17). These parameters include the LD parameter
and the durations of the applied laser pulses. Firstly, the
requisite LD parameter and the duration of the off-resonant  
red-sideband pulse for realizing the $CZ$ gate are determined by the equation (16), 
which implies
\begin{equation}
\frac{\Omega_{1,k+1}}{\Omega_{0,k}}=\sqrt{k+1}-\frac{\eta^2}{\sqrt{k+1}}
=\frac{q-0.5}{p},\hspace{3mm} t=2p\pi/\Omega_{0,k}=T_z,\hspace{3mm} p,q=1,2,3,...,
\end{equation}
Secondly, using the LD parameter determined from the above equation,
the durations of two resonant pulses 
surrounding the $k$th red-sideband pulse should be determined by solving the equation (21) 
for their different initial phases, e.g.,
\begin{eqnarray}
\left\{
\begin{array}{l}
t_1=\pi(\frac{p'+1}{\Omega_{0,0}}+\frac{q'+0.25}{\Omega_{1,1}}),\hspace{0.2cm}
t_3=\pi(\frac{p'+1}{\Omega_{0,0}}-\frac{q'+0.25}{\Omega_{1,1}}),
\hspace{0.2cm}{\rm for}\hspace{0.2cm}\phi_1=\phi_3=\pi/2;\\
\\
t_1=\pi(\frac{p'+1}{\Omega_{0,0}}-\frac{q'+0.25}{\Omega_{1,1}}),\hspace{0.2cm}
t_3=\pi(\frac{p'+1}{\Omega_{0,0}}+\frac{q'+0.25}{\Omega_{1,1}}),
\hspace{0.2cm}{\rm for}\hspace{0.2cm}\phi_1=\phi_3=3\pi/2.
\end{array}
\right.
\end{eqnarray}
Obviously, these parameters mightily are dependent of the frequency of the applied red-sideband laser light i.e., the choice of $k$, once the atomic transition frequency and trap 
frequency are defined. For different $k$ values, 
some values of these experimental parameters for this realization are given in Table I by numerical method.
\begin{center}
TABLE. some experimental parameters for realizing the exact CN gate by 
three-step sequential pulses with $\phi_1=\phi_3=\pi/2$.  
\end{center}
\begin{center}
\begin{tabular}{|c|c|c|c|c|c|c|c|c|}
\hline
$k$ & $\eta$  &$\Omega t_2/\pi$ &$\Omega t_1/\pi $&$\Omega t_3/\pi $
    & $\eta$  &$\Omega t_2/\pi$ &$\Omega t_1/\pi $&$\Omega t_3/\pi $ \\ \cline{1-9}
\hline
$1$ &$0.96920$&         $13.202$&$29.179$         &$2.8108$ 
    &$0.56625$&         $66.340$&$3.2117$         &$1.4838$ \\  \cline{4-5}\cline{8-9}
	
	&         &         $39.607$&$32.377$         &$6.0098$    
	&         &         $\cdots$&$9.0152$         &$3.7584$ \\ \cline{4-5}\cline{8-9}
	    
	&         &         $66.012$&$35.576$         &$9.2087$ 
	&         &         $      $&$13.711$         &$5.0713$ \\  \cline{4-5}\cline{8-9}
	
	&         &         $118.82$&$38.775$         &$12.408$  
	&         &         $      $&$19.514$         &$3.9634$ \\  \cline{4-5}\cline{8-9}
	 
    &         &         $\cdots$&$41.974$         &$15.607$  
	&         &         $      $&$24.211$         &$8.6589$\\  \cline{4-5}\cline{8-9}
	
	&         &                 &$45.173$         &$18.805$   
	&         &                 &$25.449$         &$16.811$\\  \cline{4-5}\cline{8-9} 
	
    &         &                 &$\cdots$         &$\cdots$ 
	&         &                 &$\cdots$         &$\cdots$\\  \cline{2-9}
    &$0.48191$&         $18.645$&$2.9777$         &$1.5148$  
	&$0.30135$&         $128.90$&$2.6684$         &$1.5174$\\  \cline{4-5}\cline{8-9}
	
    &         &         $55.934$&$5.2239$         &$3.7611$
	&         &         $\cdots$&$6.8542$         &$5.7032$\\  \cline{4-5}\cline{8-9}
	
    &         &         $\cdots$&$8.1496$         &$8.3539$
	&         &                 &$15.853$         &$8.9029$\\  \cline{4-5}\cline{8-9}
	
    &         &                 &$13.322$         &$15.594$
	&         &                 &$19.621$         &$13.866$\\  \cline{4-5}\cline{8-9}
	 
    &         &                 &$19.381$         &$12.067$
	&         &                 &$21.923$         &$11.564$\\  \cline{4-5}\cline{8-9}
	 
    &         &                 &$\cdots$         &$\cdots$  
	&         &                 &$\cdots$         &$\cdots$\\  \cline{2-9}
	&$0.23549$&         $69.853$&$2.6005$         &$1.5119$ 
	&$0.17379$&         $327.14$&$2.5538$         &$1.5070$\\  \cline{4-5}\cline{8-9}
	 
    &         &         $\cdots$&$4.6567$         &$3.5678$
	&         &         $\cdots$&$6.6147$         &$5.5679$ \\  \cline{4-5}\cline{8-9}
	 
    &         &                 &$6.8337$         &$1.3913$
	&         &                 &$12.706$         &$11.659$ \\  \cline{4-5}\cline{8-9}
	
    &         &                 &$8.7692$         &$7.6807$
	&         &                 &$16.831$         &$11.596$\\  \cline{4-5}\cline{8-9}
	
    &         &                 &$12.882$         &$11.793$
	&         &                 &$20.954$         &$11.533$\\  \cline{4-5}\cline{8-9}
	 
    &         &                 &$\cdots$         &$\cdots$
	&         &                 &$\cdots$         &$\cdots$ \\  \cline{2-9}
	
	&$\cdots$ &         $\cdots$&$\cdots$         &$\cdots$
	&$\cdots$ &         $\cdots$&$\cdots$         &$\cdots$ \\  \cline{1-9}         
\hline
$2$ &$0.78641$&         $49.846$&$4.5099$         &$0.9395$  
    &$0.50753$&         $61.714$&$3.0410$         &$1.5089$\\ \cline{4-5}\cline{8-9}
	
    &         &         $\cdots$&$9.9593$         &$6.3889$
	&         &         $\cdots$&$7.5908$         &$6.0587$\\ \cline{4-5}\cline{8-9}
	
	&         &                 &$18.133$         &$14.563$ 
	&         &                 &$12.141$         &$10.608$\\ \cline{4-5}\cline{8-9}
	
	&         &                 &$23.583$         &$20.012$
	&         &                 &$14.416$         &$12.883$\\ \cline{4-5}\cline{8-9}
	
    &         &                 &$33.448$         &$15.596$
	&         &                 &$16.690$         &$15.158$\\ \cline{4-5}\cline{8-9}
	
    &         &                 &$43.314$         &$11.180$
	&         &                 &$22.030$         &$14.369$ \\ \cline{4-5}\cline{8-9}
	 
    &         &                 &$\cdots$         &$\cdots$
	&         &                 &$\cdots$         &$\cdots$\\ \cline{2-9}
	
    &$0.27778$&         $609.54$&$2.6418$         &$1.5156$
	&$0.81347$&         $119.01$&$4.8421$         &$0.72655$\\ \cline{4-5}\cline{8-9}
	     
	&         &         $\cdots$&$6.9729$         &$1.3417$
	&         &         $\cdots$&$10.411$         &$6.2952$\\ \cline{4-5}\cline{8-9}
	
    &         &                 &$8.8778$         &$7.7516$
	&         &                 &$13.195$         &$9.0796$  \\ \cline{4-5}\cline{8-9}
	
    &         &                 &$13.035$         &$11.909$   
	&         &                 &$18.764$         &$14.648$\\ \cline{4-5}\cline{8-9}
	
    &         &                 &$15.114$         &$13.988$  
	&         &                 &$21.548$         &$17.433$\\ \cline{4-5}\cline{8-9}
		
    &         &                 &$\cdots$         &$\cdots$   
	&         &                 &$\cdots$         &$\cdots$\\ \cline{2-9}
	   
	&$\cdots$ &         $\cdots$&$\cdots$         &$\cdots$
	&$\cdots$ &         $\cdots$&$\cdots$         &$\cdots$\\ \cline{1-9}
\hline
$3$ &$0.70711$&         $71.168$&$3.8521$         &$1.2840$  
    &$0.40825$&         $939.48$&$2.8260$         &$1.5217$\\ \cline{4-5}\cline{8-9}
	
    &         &         $355.84$&$6.4201$         &$3.8521$
	&         &         $\cdots$&$7.1736$         &$5.8693$\\ \cline{4-5}\cline{8-9}
	
	&         &         $\cdots$&$14.124$         &$11.556$ 
	&         &                 &$11.521$         &$10.217$\\ \cline{4-5}\cline{8-9}
		 
    &         &                 &$\cdots$         &$\cdots$
	&         &                 &$\cdots$         &$\cdots$\\ \cline{2-9}
	
	&$\cdots$ &         $\cdots$&$\cdots$         &$\cdots$
	&$\cdots$ &         $\cdots$&$\cdots$         &$\cdots$\\ \cline{1-9}
\hline
\end{tabular}
\end{center}
The durations of the applied pulses are given by the quantities
$\Omega t_{j}/\pi, j=1,2,3$ in the table. It is seen from the table
that the switching speed of the $CN$ gate depends on the Rabi frequency
$\Omega $ and the LD parameter $\eta$. It may be estimated numerically
For example,
in the case of of a recent experiment \cite{Mor1} a single $^{9}$Be$^{+}$ ion
confined in a coaxial-resonator radio frequency (RF)-ion trap and
cooled to its quantum ground state by Ramman cooling, the target qubit
is spanned by two $^{2}S_{1/2}$ hyperfine ground states of $^{9}$Be$^{+}$:
$%
^{2}S_{1/2}|F=2,m_{F}=2\rangle$ and $^{2}S_{1/2}|F=2,m_{F}=1\rangle$,
separated by 
$\omega
_{0}/{2\pi }=1.25$\,GHz. If the Rabi frequencies are chosen as \cite
{Mor1} $\Omega =2\pi \times 140$\,kHz for resonant excitations and $\eta
\Omega =2\pi \times 30$\,kHz for off-resonant excitations, one can easily
show that the shortest duration of the applied pulses for the realization
of the exact $CN$ gate is about $10^{-4}$\,sec. (10\,kHz order). 
We see also that the durations of the first 
and third (resonant) pulses are less than the second (off-resonant) one. So 
the switching speed of the $CN$ gate is mainly dependent of the speed of the 
two-qubit $CZ$ gate operation, which is approximately $10kHz$ order. This speed
is the same as that of $CN$ gate demonstrated in \cite{Mor1}.

\section{Conclusions and discussions}
So far we have proposed a theoretical scheme to exactly implement several typical
unitary operations 
on the internal state of a single trapped cold ion e.g. the Hadamard ($H$), controlled-Z 
($CZ$) and controlled-NOT ($CN$) or controlled-X ($CX$) gates by 
using the basic one-qubit operation $\hat{r}_m(\phi,t)$ and two-qubit joint 
operation $\hat{R}_m(\phi,t)$. 
Once the coupling parameter $\eta $ is set up appropriately, we have shown that: 

1). the $CZ$ logic operation can be implemented by using a single off-resonant pulse with duration $T_z$, 

2). the $H$ gate operation can be realized by two-step sequential pulse:
an off-resonant pulse and a resonant pulse with a defined initial phase, especialy, 

3). an off-resoant pulse (for realizing the $CZ$ gate) surrounded 
by two resonant pulses (for simply rotating the state of the target) yields to 
the exact $CN$ logic operation.\\
Of course, the relevant durations of the applied pulses should be set up accurately 
also for these realizations. 

Compared to other methods \cite{Mor1}\cite{Mor2}\cite{LG} for implementing
the quantum logic operations with a single trapped ion, we emphasize that the present 
scheme has some advantages. First, although the same number of pulses are needed for 
realizing the $CN$ gate as that reported in \cite{Mor1}, the auxiliary atomic level 
and LD approximation are not required in our scheme. Consequently, pluses with different 
polarizations are not needed and our scheme can operates beyond LD limit. Second, the controlled operation
constructed in the present paper is the exact $CN$ gate, while the controlled operation
implemented in \cite{Mor2}\cite{LG} is not the exact $CN$ operation, although 
it is equivalent to the exact $CN$ gate under local transformation. Subsequent pulses
must be carried out in order to eliminate the additional phase factors \cite{Mor2}. 
So, our scheme provides an alternative method differing from the previous ones for 
realizing the exact quantum logic operations 
with a single trapped two-level cold ion. 
We hope it may be demonstrated in future experiment. 

\section*{acknowledgments}

This work was supported by the Young Science Foundation of Shanghai, China and the Natural Science Foundation of 
China.

\end{document}